\documentclass[12pt]{article}
\usepackage{amsmath}
\usepackage{graphicx,psfrag,epsf}
\usepackage{enumerate}
\usepackage{natbib}
\usepackage{url} 
\usepackage{fancyvrb}

\newcommand{\blind}{0}

\begin{document}

\def\spacingset#1{\renewcommand{\baselinestretch}%
{#1}\small\normalsize} \spacingset{1}


\if0\blind
{
  \title{\bf A Framework for Infusing Authentic Data Experiences Within Statistics Courses}
  \author{Scott D.\ Grimshaw \thanks{
    The author gratefully acknowledges Nicholas J.\ Horton for his leadership of the working group that developed the 2014 Guidelines, Natalie J.\ Blades, Chris Dixon, students in the Fall 2014 BYU Stat 330 course, Emily Juchau, and the Editor, Guest Editor, and Referees whose comments have greatly improved this paper.}\hspace{.2cm}\\
    Department of Statistics, Brigham Young University}
  \maketitle
} \fi
\if1\blind
{
  \bigskip
  \bigskip
  \bigskip
  \begin{center}
    {\LARGE\bf A Framework for Infusing Authentic Data Experiences Within Statistics Courses}
\end{center}
  \medskip
} \fi
\bigskip

\begin{abstract}
\noindent
Working with complex data is one of the important updates to the 2014 ASA Curriculum Guidelines for Undergraduate Programs in Statistical Science. Infusing `authentic data experiences' within courses allow  students opportunities to learn and practice data skills as they prepare a dataset for analysis. While more modest in scope than a senior-level culminating experience, authentic data experiences provide an opportunity to demonstrate connections between data skills and statistical skills. 
The result is more practice of data skills for undergraduate statisticians.
\end{abstract}

\noindent%
{\it Keywords:}  assessment, undergraduate statistics education, curriculum guidelines
\vfill

\newpage

\spacingset{1.45} 

\section{Introduction}
\label{sec:intro}

The 2014 ASA Curriculum Guidelines for Undergraduate Programs in Statistical Science (\cite{Guidelines} and hereafter referred to as ``2014 Guidelines'') define the skills needed for statistics majors. Commonly, statistics programs are designed to include courses in each skill area, followed by a senior-level capstone, internship, or research experience. The `culminating experience' is important for students because statistics contains many connections between statistical application, statistical theory, data manipulation, computation, mathematics, and communication.

Excellent capstone courses for statistics majors have been shared in the literature (see \cite{Capstone1} and \cite{Zhu2013}), but the growth in undergraduate statistics programs indicates that capstones need to be scaled up from small courses with one-on-one interactions with a professional statistician to courses with enrollments of 50 to 100 students each year without losing the characteristics that make a capstone experience valuable. Even in programs with strong capstone experiences, the 2014 Guidelines point out that students should have a scaffolded exposure to topics and connections throughout the academic program, rather than relying on a single senior-level course to tie everything together.

Connections between methods, theory, data, and mathematics are easily lost since undergraduate programs offer multi-course sequences in these topics over as many as four years. Integration also suffers when faculty teach only one course in the curriculum, which is often in the area of their greatest interest. Some students fail to develop connections between topics.
An important aspect of program assessment is evaluating the presence and frequency of information silos.

The integration of some topics in the undergraduate program is common. For example, most programs integrate mathematics foundations with statistical theory and integrate real-world data with statistical methods. Other topics are harder to integrate in the curriculum but have been successfully demonstrated in the statistical education literature. For example, \cite{Horton2013} tackled integrating statistical theory and computation by pointing out the pedagogical benefits and providing example projects.

An authentic data experience provides data sources and data instructions as part of an analysis. The goal is for students to increase their opportunities to manipulate and restructure data that are provided in different formats. 
Creating real data applications is a natural part of course preparation, but teachers often perform the data collection and management tasks and then provide students with a curated data file. 
For example, one of the strengths of \cite{Sheather} as a regression textbook is the many applications which required significant data acquisition and manipulation work by the author, but all the datasets are presented to students as clean, well-organized files, easily read from the book website.
Using the vocabulary of \cite{Wickham}, teachers hide the `messy data' aspects and provide `tidy data' --- even when students possess the data skills required to work with the messy data. 
It is valuable for students to not only have many authentic data experiences but also to have the professor model the correct application of statistics by showing work with messy data in lectures.

While authentic data experiences in capstone courses and DataFest (see \cite{DataFest1} and \cite{DataFest2}) provide opportunities to work with messy data, more can and should be done in other courses to allow students to practice problem-solving using data.
Authentic data experience should supplement not avoid creating and offering courses to satisfy the data manipulation and computation skills in the 2014 Guidelines and the incorporation of data science courses into the curriculum. 

What is good for statistics majors can also be applied to introductory courses. There may be no data skills on the learning outcomes for introductory courses, but \cite{HortonCHANCE} advocate for complex and interesting data in such courses.
Some examples and homework in an introductory course may be modified and/or updated to use the original source data instead of a curated dataset. The data skills required would certainly be modest and need to fit student backgrounds. The objective would be for students to see that data skills are required in an analysis. Students may rely on code provided to them that results in their own copy of the dataset.   

The paper proceeds with proposed metrics to evaluate authentic data experiences for lectures, homework, and exams in Section \ref{sec:metrics}. Section \ref{sec:examples} describes three examples of authentic data experiences used in teaching a regression course for statistics majors. The paper closes with an assessment of the infusion of authentic data experiences in a BYU Fall 2014 class followed by a summary and discussion.

\section{Metrics to Evaluate Authentic Data Experiences}
\label{sec:metrics}

Statistics education is rich with applications that satisfy the definition of `real data' in the GAISE college report (\cite{GAISE}) and are available in textbooks, journals (e.g.\ {\it JSE} Data Sets and Stories), and internet repositories (e.g.\ StatLib, Data and Story Library, R libraries). A common expectation with  data sets is the `story' --- shared with students --- that motivates the research questions that led to data collection. Student learning is enhanced if the research story also has a data story. 

Authentic data experiences specifically include details and instructions for any tasks requiring two  skills mentioned in the 2014 Guidelines: (1) the ``ability to manage and restructure data'' and (2) ``data manipulation using software in a well-documented and reproducible way, data processing in different formats, and methods for addressing missing data.'' 

In order to distinguish between the breadth and depth of skills required for an authentic data experience, two dimensions for breadth are proposed: ``Data from Different Sources and Formats'' and ``Data Manipulation.'' Within these dimensions a `good/better/best' classification is suggested as a rubric to distinguish depth of skills, and Table \ref{table1} defines tasks required on the data step of the analysis. Another way to think of the classification is the time expected for students to create the dataset.

\begin{table}
\caption{Breadth and Depth of Data Skills}\label{table1}
\begin{center}
\begin{tabular}{| l p{4.5in} |}
\multicolumn{2}{l}{{\bf Data from Different Sources and Formats}} \\
\hline
{\it Good:} &
Read a space-, tab-, or comma-delimited file with or without a first header row of variable names \\
{\it Better:} &
Read data from an HTML table or other complex format \\
{\it Best:} &
Read data from multiple sources or multiple HTML tables \\
\hline
\multicolumn{2}{c}{\ } \\
\multicolumn{2}{c}{\ } \\
\multicolumn{2}{l}{{\bf Data Manipulation}} \\
\hline
{\it Good:} &
Conventions for interpretable variable names \\
{\it Better:} &
Compute additional variables and/or subset the data \\
{\it Best:} &
Merge or combine data from multiple sources to create dataset for analysis and/or identify missing and unreliable observations and/or variables \\
\hline
\end{tabular}
\end{center}
\end{table}

To demonstrate the classification, consider the climate science example of \cite{Witt2013} to teach simple linear regression. One could use the curated data provided as supplemental material at \url{http://www.amstat.org/publications/jse/v21n1/witt/temp_co2_data.txt}, and doing so would be `good' for both dimensions since that is a clean, well-organized, easily read file with a header row of interpretable variable names. As an alternative, students could be directed to the different webpages containing the original data that are space-delimited files with interpretable variable names and instructed to merge the datasets by year to form the dataset for analysis. A small change to include data skill tasks enriches the teaching experience to `best' on both dimensions because data are read from multiple sources and merged.

In order to classify or assess the data skills required for a particular authentic data experience, two  important topics must be shared: Data Details and Teaching Notes. The Data Details section should include the data source, characteristics of the data, and the instructions to construct the dataset for analysis. The Teaching Notes should address how classes with different data skill backgrounds or prerequisites may need  instructions or code elements in order to perform the required tasks. To demonstrate on the climate science example of \cite{Witt2013} the Data Details would include that the source of the Annual Temperature Data is \url{http://data.giss.nasa.gov/gistemp/graphs_v3/Fig.A2.txt} and point out the following features when accessed 31 March 2015:

\begin{itemize}
\item
space-delimited text file
\item
4 lines of text before the data
\item
column header on line 3, data begins on line 5
\item
the current year is the last row but is coded as missing with {\tt *}
\item
attempt to read file directly from source in R with read.table results in error:
\begin{Verbatim}[frame=single,baselinestretch=1.0,fontsize=\small]
Error in file(file, "rt") : cannot open the connection
In addition: Warning message:
In file(file, "rt") : cannot open: HTTP status was '403 Forbidden'
\end{Verbatim}
\end{itemize}
The source for the Annual ${\rm CO}_2$ Data is \url{ftp://aftp.cmdl.noaa.gov/products/trends/co2/co2_annmean_mlo.txt}
and points out the following features when accessed 31 March 2015:

\begin{itemize}
\item
space-delimited text file
\item
can be read directly from source
\item
50 lines of text describing the data before the data (each line begins with {\tt \#})
\item
column header line begins with {\tt \#}
\end{itemize}

One example of the Teaching Notes is using the climate science example in lecture of a regression course for statistics majors as an active learning activity. After providing the Data Story from \cite{Witt2013} and the Data Details section, I ask students to graph and estimate a model for the relationship between ${\rm CO}_2$ and temperature. I have them discover the data complications instead of pointing them out, but one could point them out to be more time efficient. Students seem to quickly understand what they want to do, but common questions involve syntax for reading from a webpage, creating headers, and skipping lines.

Some of the students perform the analysis in SAS; most of these students use {\tt datalines}, where they copy and paste from the two source webpages. The {\tt proc merge} is basic, and students who haven't learned that topic follow the syntax when it is provided.
Some of the students perform the analysis in R, where the ${\rm CO}_2$ data can be read from the source with {\tt read.table}, \verb|read_table| from the {\tt readr} library, or {\tt fread} from the {\tt data.table} library, but the same is not true for the temperature data because of the error message. The syntax for creating descriptive variable names 
should be discussed, and some students are not familiar with the syntax for skipping lines in the file.
Because the last row of the temperature data is missing and identified as {\tt *}, the R default is to make the column a factor. There are no complications to using the {\tt merge} function.
Some students will combine and merge the data using Excel. A comparison of student approaches in SAS, R, and Excel provides an opportunity to talk about reproducibility and documenting code.

While not part of the metrics to evaluate authentic data experiences, one feature of articulating the required data skills is that research questions can be addressed using currently available data. For example,  \cite{Albert} provides curated seasonal batting data from 1871--2009 at \url{http://www.amstat.org/publications/jse/v18n3/mlb_batting.dat.txt}. 
With the passage of time, the examples of Derek Jeter and Alex Rodriguez as current players and batting trends up to 2009 are less timely to students interested in sports, and what was once a current research question has become stagnant and ossified. However, \cite{Albert} provides sufficient detail about the collection of files on the Lahman Baseball Database and tasks required to filter and merge the master and batting files that the research questions and teaching notes can be updated to reflect current players and trends. The {\it JSE} Data Contributors Guidelines consider non-static datasets when the teaching notes include annotations and commentary in the form of reproducible code such as \cite{Albert} to make the dataset applicable to a large audience. 

\section{Examples}
\label{sec:examples}

Fall 2014 was a new preparation for a regression course for statistics majors and provided the opportunity to incorporate authentic data experiences into lectures, homework, and exams. The course is a traditional undergraduate treatment of linear regression, logistic regression, and time series that also satisfies the Society for Actuaries Applied Statistics VEE (Validation by Educational Experience). The course prerequisite is a `second course in statistics' based on designed experiments, and the required textbook Fall 2014 was \cite{Sheather}, but \cite{Weisberg} is also at the appropriate level. 

Table \ref{table2} summarizes the 55 different real data applications used in lecture examples, homework assignments, and exams assessed by the metrics proposed in Section \ref{sec:metrics}. One stand-out table entry shows that most (38/55) of the applications are in files that are comma- and space-delimited, either with or without a header row. Further investigation reveals that 15 of these 38 applications were in lectures, where using applications from the textbook reinforced reading assignments, and 13 were in exams or practice exams, where the exam questions were on the regression material. Most of the applications corresponding to `better' and `best' assessments are homework assignments, but it is important to teach and demonstrate the expected data acquisition and manipulation skills during lectures.

\begin{table}
\caption{Authentic Data Experiences in BYU Stat 330 Fall 2014}\label{table2}
\begin{center}
\begin{tabular}{l c | r r r }  
\multicolumn{2}{c}{} 
     & \multicolumn{1}{p{1.6cm}}{} & \multicolumn{1}{p{1.6cm}}{} & \multicolumn{1}{p{1.6cm}}{} \\
 & & \multicolumn{3}{c}{Different Sources \& Formats} \\
 & & \multicolumn{1}{c}{Good} & \multicolumn{1}{c}{Better} & \multicolumn{1}{c}{Best} \\
\hline
Manipulation 
 & Good   & 38 & 1 & 0 \\
 & Better &  2 & 9 & 0 \\
 & Best   &  0 & 2 & 3 \\
\hline
\end{tabular}
\end{center}
\end{table} 

One of the challenges of writing authentic data experiences is identifying applications with some, but not significant data acquisition and manipulation tasks and providing clear instructions for students in the Data Details section regarding what tasks they need to complete to prepare the dataset for the analysis questions on the assignment. It is important to remember that practicing data-related skills is secondary in most courses, and a rule of thumb is that students spent 15 to 20 minutes preparing the dataset for analysis. 
While this restriction would not be needed if the applications were used in a data course, the Teaching Notes often include code chunks for tasks the students may not know how to perform or which would take them longer than 15 to 20 minutes to learn, write and debug.
Providing code chunks is one response to the challenge in \cite{Cobb} to flatten the prerequisites.
It may appear the three applications that are `best' in both dimensions would be ideal, but in fact two resulted in students complaining about excessive time to create the dataset:
one example required downloading 12 files from queries to the US NOAA website with students reporting it took 40 minutes to create the dataset, and the other example required deeper knowledge of NBA rosters than most students found valuable to research.  
Only the climate science application from \cite{Witt2013} fell within the 15 to 20 minutes to prepare the dataset.

Three examples are provided to demonstrate authentic data experiences for homework assignments and provide a template for creating future authentic data experiences. The Data Story and Data Details sections and Teaching Notes are provided for each example, as well as the classification of the breadth and depth of data skills. While all three examples are from one course for statistics majors, the intention is to demonstrate authentic data experiences throughout the statistics curriculum.

\subsubsection*{Difference Between MLB Leagues and Divisions}

Different Sources and Formats: Best (read data from multiple HTML tables, create additional variables from table layout) \\
Manipulation: Better (filtering rows and columns from the dataset) 

\noindent
{\sc Data Story:}
MLB organizes the 30 teams into two leagues (American League, National League) with three divisions based on geography (East, Central, West) for each league. Unlike other professional leagues, MLB has different rules for each league that has led to long and passionate arguments between fans of the two leagues (Google `Designated Hitter' for more information). Each year there are also arguments that particular divisions are stronger or weaker in terms of what it takes to win.
Consider a model where the response variable is the number of wins in a season (\texttt{wins}), and the explanatory variables are:
\begin{itemize}
\item
Factor \texttt{league} with levels \texttt{AL} for American League and \texttt{NL} for National League
\item
Factor \texttt{division} with levels \texttt{East}, \texttt{Central}, \texttt{West}
\item
\texttt{run.diff}, the run differential for the season (difference between runs scored and runs allowed)
\end{itemize}

\noindent
{\sc Data Details:}
The number of wins and the run differential for each team are reported on MLB Standing pages on many sports websites --- usually as HTML tables. The league and division for each team is created from the structure of the MLB Standings webpage.
What follows is when the data source is \url{ESPN.com} and accessed 31 March 2015:
\begin{itemize}
\item
The webpage \url{http://espn.go.com/mlb/standings} provide the current standings. During the season the standings change daily. During spring training the standings are organized by Cactus and Grapefruit League (teams having spring training in Arizona and Florida, respectively) and not the factor levels described above. The web address for previous season's standings are found by changing the ``Season:'' variable in the table. For example, the 2014 MLB Regular Season is at 
\url{http://espn.go.com/mlb/standings/_/season/2014}
\item
The standings are in two HTML tables on the same webpage
\item
The standings have both numeric and character variable types
\item
The standings contain rows and columns that should be ignored
\item
Create League and Division variables from the structure of the standings
\end{itemize}

\noindent
Summary of Data Skills: 
\begin{itemize}
\item
Reading data from an HTML table
\item
Reading numeric and character variables
\item
Filtering rows 
\item
Filtering columns
\item
Creating additional variables from table layout
\end{itemize}

\noindent
{\sc Teaching Notes: Multiple Regression with Qualitative and Quantitative Explanatory Variables in a Course for Statistics Majors} 

\noindent
My objective is for students to write reproducible code in R to create a dataset from two HTML tables. The example is assigned in homework. The following notes are provided to help the students to write the code:
\begin{itemize}
\item
Use the function \texttt{readHTMLTable} from the \texttt{XML} R library. 
\item
Use the \texttt{which=1} and \texttt{which=2} declaration to read the 1st table (American League) and 2nd table (National League) from the webpage.
\item
Use the \texttt{colClasses} argument to specify the `type' of each column as \texttt{"numeric"} for numerical data columns and \texttt{"character"} for character data columns.
\item
Use \texttt{rbind} to combine into a single dataset.
\item
Use \texttt{header=FALSE} and specify your own column names. 
\item
Use \texttt{skip.rows} to only read the 30 teams (ignoring all other rows). 
\item
Use the organization of the webpage table to create \texttt{league} and \texttt{division}.
\end{itemize}
\begin{Verbatim}[frame=single,baselinestretch=1.0,fontsize=\small]
Solution (not provided to the students)

# Is There a Difference between MLB Leagues and Divisions 
# (after adjusting for scoring)?
# Source Data 2014 MLB Season Accessed 31 March 2015

# read 2014 data from webpage
library(XML)

# AL is Table 1
al<-readHTMLTable("http://espn.go.com/mlb/standings/_/season/2014",
                  which=1,header=FALSE,skip.rows=c(1,7,13),
                  colClasses=c("character",rep("numeric",11)))
# note: this use of colClasses will leave some columns with missing values

# NL is Table 2
nl<-readHTMLTable("http://espn.go.com/mlb/standings/_/season/2014",
                  which=2,header=FALSE,skip.rows=c(1,7,13),
                  colClasses=c("character",rep("numeric",11)))

# combine AL and NL
mlb<-rbind(al,nl)

# subset columns to only those of interest
mlb<-mlb[,c(1,2,10)]

# assign column names
names(mlb)<-c("team","wins","run.diff")

# identify leagues
mlb$league<-c(rep("AL",15),rep("NL",15))

# identify divisions
mlb$division<-rep(c(rep("East",5),rep("Central",5),rep("West",5)),2)
\end{Verbatim}
The first question in the homework is ``write a webscraper.'' The full homework assignment is at \url{http://grimshawville.byu.edu/hwMLBfromHTML.pdf}. The estimated time for students to create the dataset is approximately 20 minutes.

\subsubsection*{Forecasting Movie Box Office Revenue}

Different Sources and Formats: Better (read data from an HTML table) \\
Manipulation: Good: (change variable name to something less confusing) 

\noindent
{\sc Data Story:}
In the early days of the movie industry, customers could only see a movie during a finite time period in a theater. How that has changed! Today movies and other entertainment content are still available in theaters but are also available to customers for longer and more flexible time periods on digital and disk formats. From a business perspective, each of the different channels (theaters, Netflix, Amazon, YouTube, iTunes, TV broadcasters, Redbox) seeks to `monetize' the content they own or license. The traditional entertainment revenue is `theater box office,' defined as the amount paid by customers to watch a movie shown in a theater. While movies have multiple revenue streams (for example, international box office, DVD sales, digital rights or downloads), the theater box office usually drives the later revenue streams and for some independent movies (like ÒNapoleon DynamiteÓ) the appearance in theaters is a mark of success. 
Focusing on the economic outlook of theater box office, consider forecasting the next five years. The response variable is the annual gross box office, defined as the total revenue from all movies seen by customers in theaters in a given calendar year.

\noindent
{\sc Data Details:}
Box Office Mojo is a website reporting on many of the business aspects of the movie industry. The webpage \url{http://www.boxofficemojo.com/yearly/} tables the gross box office for all movies in theaters in a calendar year (in \$ million).  
\begin{itemize}
\item
The tabs on the webpage are actually the first HTML table, so the data of interest is the second HTML table
\item
The table contains 10 columns with different formats: numeric, accounting (leading \$, commas at thousands), percentage (trailing \%), character
\item
As is common on webpages the most current data is at the top of the page. Time series data usually begins $t=1$ (oldest data) to $t=T$ (newest data).
\item
The first row is the current year to date gross box office. The other values are for 365 calendar days.
\end{itemize}

\noindent
Summary of Data Skills:
\begin{itemize}
\item
Reading data from an HTML table
\item
Reading variables with multiple formats
\item
Change the order from ``most recent data first'' to time series
\item
Remove current year since the value is YTD
\end{itemize}

\noindent
{\sc Teaching Notes: Time Series ARIMA Model and Forecasts in a Course for Statistics Majors} 

\noindent
My objective is for students to write reproducible code in R to create a dataset from the HTML table. The example is assigned in homework. The following notes are provided to help the students to write the code:
\begin{itemize}
\item
Use the function \texttt{readHTMLTable} from the \texttt{XML} R library. 
\item
Use the \texttt{which=2} declaration to read the second HTML table from the webpage. The first HTML table corresponds to the tabs on the webpage.
\item
Use the \texttt{colClasses} argument to specify the `type' of each column as \texttt{"numeric"} for numerical data columns, \texttt{"character"} for character data columns, \texttt{"FormattedNumber"} for numerical data columns with `,' separating thousands. There is no R defined format for \$\texttt{dd,ddd.dd} but the following code creates an \texttt{AccountingNumber} `type' that can be run before the \texttt{readHTMLTable} code and then used in the \texttt{colClasses} specification:
\begin{Verbatim}[frame=single,baselinestretch=1.0,fontsize=\small]
# no predefined format for $dd,ddd.dd create one called "AccountingNumber"
setClass("AccountingNumber")
setAs("character", "AccountingNumber", 
      function(from) as.numeric(gsub(",", "", gsub("[:$:]", "", from))))
\end{Verbatim}
\item
Use \texttt{header=TRUE} but then change the variable name for gross box office (\$ million) from \texttt{TotalGross*} to \texttt{Gross} since the `\texttt{*}' is confusing. 
\item
Remove the row for the current year since it is YTD.
\item
Reorder the data to comply with the time series convention of oldest data (first row) to current data (last row).
\end{itemize}
\begin{Verbatim}[frame=single,baselinestretch=1.0,fontsize=\small]
Solution (not provided to the students)

# Box Office Mojo (IMDb) 
# Yearly Total Gross ($ millions) Box Office (1980 to present)
# http://www.boxofficemojo.com/yearly/

# multiple column formats
# while it looks like only one HTML table on website, the tabs are 
# actually the first HTML table and the data of interest is 
# in the second HTML table

# use webscraper to read the webpage HTML table into R
library("XML")

# no predefined format for $dd,ddd.dd so create one called "Accounting Number"
setClass("AccountingNumber")
setAs("character", "AccountingNumber", 
      function(from) as.numeric(gsub(",", "", gsub("[:$:]", "", from))))

# read data from HTML table
boxoffice<-readHTMLTable("http://www.boxofficemojo.com/yearly/",
                         which=2,header=TRUE,skip=1,
                         colClasses=c("numeric","AccountingNumber","Percent",
                                      "FormattedNumber","Percent",
                                      "numeric","FormattedNumber",
                                      "AccountingNumber","AccountingNumber",
                                      "character"))
# note: ignore warning messages about creating NAs 
#       (the HTML table contains missing values)

# subset columns to only those of interest (year and gross) 
boxoffice<-boxoffice[,1:2]

# create interpretable variable name (from HTML table TotalGross*)
names(boxoffice)[2]<-"Gross"

# remove current year (first row) since it doesn't represent 365 days
boxoffice<-boxoffice[-1,]

# table is ordered from newest to oldest, so restructure 
boxoffice<-boxoffice[order(boxoffice$Year),]
\end{Verbatim}
The first question is ``write a webscraper.'' The full homework assignment uses the \texttt{astsa} R library to estimate and forecast an ARIMA(1,1,1) model and is at \url{http://grimshawville.byu.edu/hwTimeSeriesHTML.pdf}.
The estimated time for students to create the dataset is approximately 10 minutes.

\subsubsection*{Effect of Age and Race on Having Health Insurance}

Different Sources and Formats: Good or Better, depending on students' background with SAS Transport files \\
Manipulation: Better (filtering rows, creating interpretable variable names), but could be Best if the SAS code to merge the files is not provided

\noindent
{\sc Data Story:}
The US CDC performs a large survey of interviews and physical examinations that assess the health and nutritional status of adults and children in the US. The most recently completed, publicly available data is the NHANES 2011-2012 Survey. You have been asked to investigate the relationship between age and race on whether or not individuals have health insurance.
Consider a model where the response variable is whether or not an individual has insurance (\texttt{insured}) and the explanatory variables are:
\begin{itemize}
\item
\texttt{age}, the age of an individual (in years)
\item
Factor \texttt{race} with levels \texttt{1}=Mexican American, \texttt{2}=Other Hispanic, \texttt{3}=Non-Hispanic White, \texttt{4}=Non-Hispanic Black, \texttt{6}=Non-Hispanic Asian, \texttt{7}=Other Race - Including Multi-Racial
\end{itemize}

\noindent
{\sc Data Details:}
As is typical for the analysis of NHANES data, merging two SAS datasets is required. For NHANES 2011-12 the downloaded datasets are SAS Transport files. The respondent's age is in the Demographics Data, which must be downloaded locally from \url{http://wwwn.cdc.gov/nchs/nhanes/2011-2012/DEMO_G.XPT}, and the insurance coverage is in one of the Questionnaire Data files, which must be downloaded locally from \url{http://wwwn.cdc.gov/nchs/nhanes/2011-2012/HIQ_G.XPT}.

\begin{itemize}
\item
SAS Transport files must be downloaded and saved locally
\item
\texttt{proc merge} should only keep a subset of the variables and only keep observations in the \verb|HIQ_G.XPT| dataset
\item
Filter out observations of those who responded \texttt{REFUSED} or \texttt{DON'T KNOW} for the health insurance question
\item
Change the NHANES variable names to something more descriptive
\item
The NHANES Tutorials provide examples for the SAS code
\end{itemize}

\noindent
Summary of Data Skills:
\begin{itemize}
\item
Download publicly available SAS Transport files
\item
Merge two SAS Transport files
\item
Create interpretable variable names
\item
Filter rows
\end{itemize}

\noindent
{\sc Teaching Notes: Logistic Regression with Qualitative and Quantitative Explanatory Variables in a Course for Statistics Majors} 

\noindent
My objective is for students to download data from NHANES and write SAS code to create a dataset. The downloaded files are SAS Transport files and so it is easiest to use SAS to create the dataset. The example is assigned in homework. The following notes and code are provided to help students who haven't taken a SAS class or don't know about SAS Transport files:
\begin{itemize}
\item
Code to merge two SAS Transport files and subset to only those in NHANES that participated in insurance questionnaire:
\begin{Verbatim}[frame=single,baselinestretch=1.0,fontsize=\small]
/* NHANES 2011-2012 Data on Insured */

/*
Download NHANES 2011-2012 Demographics Data File
Name: Demographic Variables and Sample Weights 
  http://wwwn.cdc.gov/nchs/nhanes/2011-2012/DEMO_G.XPT 

Download NHANES 2011-2012 Questionnaire Data File
Name: Health Insurance 
  http://wwwn.cdc.gov/nchs/nhanes/2011-2012/HIQ_G.XPT
*/

**** note: your folder location will differ!;
libname demo  xport "E:/Classes/S330/homework/DEMO_G.XPT";
libname insur xport "E:/Classes/S330/homework/HIQ_G.XPT";

/* merge demographic data and insurance survey data 
    Age  (ridageyr)
    Race (ridreth3)
         Code Description
         1    Mexican American
         2    Other Hispanic	
         3    Non-Hispanic White	
         4    Non-Hispanic Black	
         6    Non-Hispanic Asian	
         7    Other Race - Including Multi-Racial
    Covered by Insurance (hiq011) 
         Code Description
         1    YES
         2    NO
         7    REFUSED
         9    DON'T KNOW;
*/    

data nhanes;
  merge demo.DEMO_G (keep=seqn ridageyr ridreth3)
        insur.HIQ_G (keep=seqn hiq011 in=a);
  by seqn;
  if a; * this subsets to only those in NHANES that 
          participated in insurance questionnaire;
run;
\end{Verbatim}
\item
Filter out observations of those who responded \texttt{REFUSED} or \texttt{DON'T KNOW} for the health insurance question
\item
Change the NHANES variable names to something more descriptive
\end{itemize}

\begin{Verbatim}[frame=single,baselinestretch=1.0,fontsize=\small]
Solution appended to the code provided above (not given to the students)

data nhanes;
  set nhanes;
  * subset to those that responded and define insured=(yes/no)
  if hiq011 in (1,2);
  * 'rename' variables so they are easier to remember;
  if hiq011=1 then insured="yes";
    else insured="no";
  age=ridageyr;
  race=ridreth3;
  drop ridageyr ridreth3 hiq011;
run;
\end{Verbatim}
The first question is ``download the appropriate files from NHANES and create the SAS dataset \texttt{nhanes}.'' The full homework assignment is at \url{http://grimshawville.byu.edu/hwNHANESinSAS.pdf}. 
The estimated time for students to create the dataset is approximately 10 minutes with the code provided, but it would be more if students were asked to write that SAS code.

\section{Assessment}
\label{sec:assess}

One of the advantages of using authentic data experiences in a course is an improvement in the quality of course projects. Students feel empowered to ask a question and then find the data, instead of identifying a curated dataset from a repository or another book and performing an analysis. In Fall 2014 the 45 students in Stat 330 were required to complete a course project, and most students worked in pairs (two groups had four members). Of the 19 projects, 14 used data from multiple sources (three sources was most common) that required merging, and 16 included additional data manipulation tasks of subsetting and computing additional variables. 

Another assessment comes from the Fall 2014 student ratings for the course. The mean response to `Materials and Activities Effective' was 1.0 higher than the department mean (eight-point scale). Since the authentic data experiences were part of every assignment, the difference indicates that students found them effective. While not specifically addressing data skills, the mean response to `Intellectual Skills Developed' was 0.6 higher than the department mean (eight-point scale) and indicates that, at the very least, the integration of data skills didn't distract from the course learning outcomes of regression modeling and time series. Four student comments appreciated the `real world situations,' and two appreciated how they began to see how the different skills in statistics `all fit together.' As a follow-up assessment, a survey was sent to all 45 students a semester later and asked about any change they'd experienced with regard to the six data skills organized under Different Sources and Format and Manipulation and defined as good, better, and best. While there is likely a response bias (those who didn't enjoy the course wouldn't respond), 15 of the 16 responses reported `Much Stronger' skills in at least one of the data skill sets, with 12 reporting `Much Stronger' skills in three or more of the six skills defined.  

There are a few drawbacks with authentic data experiences. The most important is what is omitted from the course when authentic data experiences are added. From the instructor perspective, new homework assignments had to be written for better and best applications since the textbook's application are all at the good level. Course preparation time shifted from some regression topics, and assessment at semester's end reflected less depth on about 15\% of the course material compared to another instructor teaching the same course. From the student perspective, using fewer datasets in the weekly homework because of the time required to assemble the data for analysis resulted in less practice on regression skills. From the Fall 2014 Stat 330 course ratings, the mean response to `Valuable Time Out of Class' was 89\%, which is higher than the department mean of 81\%. Of the 24 student comments, five were about homework load with two reporting `the right amount,' three reporting it was `too much,' and two suggesting more time in class with data sets similar to what will be assigned on the week's homework.

\section{Conclusion}
\label{sec:conc}

Statistics is a field of study requiring an integrated set of skills. The 2014 Guidelines define the expected skills of a statistics major and encourage programs to provide opportunities for students to connect these skills. `Authentic data experiences' are proposed that extend the definition of real data to include the application of data skills. 

Presently, instructors can choose from an exploding set of real data applications in their courses. Unfortunately, the end result from the student's perspective is often a clean, well-organized, easily read file. A small paradigm change is to provide source data locations with sufficiently detailed instructions for the students to prepare the dataset for analysis. 
An added benefit of finding and writing authentic data experiences is that it forces instructors out of the comfort zones described by \cite{Horton2015} and keeps instructors current as preferred technologies change over time.
Applications that are time sensitive benefit from providing the instructions to obtain current data. In time series in particular, a student analyzing data that was current at book publication may feel the application is artificial. In some cases obtaining current data for the same problem requires basic data-related skills.

Open questions include the number of real data applications that should be included in a course and program, the number that should be at the `better' and `best' levels, and the value of including `better' or `best' applications on exams in non-data courses. Further extensions are creating an environment to share applications where the teaching notes articulate how different expectations for the data skills modify the example. For example, the same application could be presented to a statistics major course and an introductory course with different student expectations with regard to data skills. Since the Internet is a dynamic environment for data, there needs to be a way for people to share when data becomes subscription-based or a website has changed.

The 2014 Guidelines challenge undergraduate statistics programs to emphasize working with complex data. Students will take courses in data manipulation and computation, as well as having a culminating experience through a capstone course, an internship, and/or a mentored research experience. The paper suggests that more can be done to develop studentsÕ skills and confidence by providing authentic data experiences in other courses.

\bigskip

\begin{center}

{\large\bf SUPPLEMENTARY MATERIAL}

\end{center}

\noindent
The three examples are available as a Wiki at \url{https://grimshaw-wiki.byu.edu}. Also available is the climate science example from \cite{Witt2013} with Teaching Notes for both a regression class and an introductory class. The Wiki welcomes contributions, comments, and updates.

\bibliographystyle{chicago}

\bibliography{Bibliography-Grimshaw1}

\end{document}